# Affective touch communication in close adult relationships

Sarah McIntyre, Athanasia Moungou, Rebecca Boehme, Peder M. Isager, Frances Lau, Ali Israr, Ellen A. Lumpkin, Freddy Abnousi, Håkan Olausson

*Abstract*— Inter-personal touch is a powerful aspect of social interaction that we expect to be particularly important for emotional communication. We studied the capacity of closely acquainted humans to signal the meaning of several word cues (e.g. gratitude, sadness) using touch sensation alone. Participants communicated all cues with above chance performance. We show that emotionally close people can accurately signal the meaning of different words through touch, and that performance is affected by the amount of contextual information available. Even with minimal context and feedback, both attention-getting and love were communicated surprisingly well. Neither the type of close relationship, nor self-reported comfort with touch significantly affected performance.

## I. INTRODUCTION

Touch is a powerful social tool, engaging many different areas of the body, and playing an important role in human development [1]–[3]. Recent evidence suggests there may be a dedicated neural network involved in signaling the pleasant, emotional, or social aspects of skin-to-skin stimulation, that is distinct from the discriminative touch function that is used for navigating and manipulating the environment [1], [4]. This importance is reflected in recent development of haptic devices concerning both communication [5] and the affective aspects of social touch [6], [7].

Despite this importance, only a few studies have explored the use of touch as a means to communicate. These studies show that unacquainted partners can robustly communicate anger, fear, disgust, love, gratitude, sympathy, happiness, and sadness [8], [9]. Although touch is often considered tertiary to auditory and visual communication, there is evidence that touch is the preferred communication channel for messages of love and sympathy [10]. Consoling or calming touch behavior is also prevalent in the parent-child relationship [2], and may persist into adult relationships. However, adult relationships vary in their degree of intimacy, and touch communication may function differently according to the strength of the emotional bond, or the nature of the relationship.

Previous studies have already shown that romantic couples are more effective at communicating through touch than strangers [11], and that loved ones and close friends touch each other on more areas of the body and for more reasons than less familiar individuals [3]. In this study, we wanted to test whether touch communication performance varies for pairs of people in different kinds of adult relationships, including romantic relationships and friendships. To do this, we measured how well adults in pre-existing relationships could communicate a variety of emotional word cues.

We also wanted to investigate the effect of contextual cues on communication performance. In previous studies of touch communication, participants typically responded from a provided list of possible emotions [8], [9], [11]. Despite the presence of an "other" option, such an approach eliminates a lot of possibilities from the receivers' interpretations of the touches. In the current study we used a similar forced-choice task design but included an open-ended free-text entry question at the beginning of the experiment for comparison. Previous studies also typically gave participants minimal emotional or social context and no feedback, which isolates the effects of touch, but likely produces worse understanding than can be achieved outside the laboratory. We also investigated whether small increases in the contextual information available could affect performance.

## II. METHODS

### A. Participants

We recruited pairs of adult participants with a pre-existing relationship in which they claimed to be comfortable touching each other in a normal social context. There were 38 participants (19 pairs), aged 19 – 68 (median = 27). Two pairs had a family relationship (1 F-M, 1 M-M), 8 were friends (1 F-F, 1 F-M, 6 M-M) and 9 were in a romantic relationship (9 F-M). Participants provided informed consent, and the study was conducted in accordance with the regulations of the regional ethics committee who approved the study.

### B. Procedure

After a brief explanation of the experimental session, participants were separated into different rooms to fill out questionnaires. The participants were then assigned initial roles, one as the toucher, the other as the receiver, and the touch communication task was explained to them. After performing the touch communication task, the participants switched roles so that the toucher became the receiver and

*Research supported by Facebook, Inc. USA

S. McIntyre, A. Moungou, R. Boehme and H. Olausson are with the Centre for Social and Affective Neuroscience, Linköping University, Linköping, Sweden (e-mail: sarah.mcintyre@liu.se).

P. M. Isager was with the Centre for Social and Affective Neuroscience, Linköping University, Linköping, Sweden, and is now with the Department of Industrial Engineering and Innovation Sciences, Eindhoven University of Technology, Eindhoven, Netherlands

F. Lau, A. Israr and F. Abnousi are with Facebook Reality Labs, USA

E. A. Lumpkin is with Colombia University, USA

vice versa, and the touch communication task was performed with the new roles. At the end of the experiment, the participants took turns to provide feedback on the touch communication task.

*C. Touch communication task*

The receiver sat in a chair with his/her left arm resting passively on an arm-rest accessible to the toucher. The rest of their body and face was obscured by a curtain (Fig. 1). On each trial, the toucher was provided with one of the toucher cues shown in Table 1. One of the words ("attention") served as a comparison word for the success rate of the communication, since we expected that it would be easily communicated. This is because getting someone's attention with touch is a common occurrence in everyday life, uses relatively stereotyped touch action, occurs among people with a wide variety of relationships, and usually no on-going social context needs to exist between those involved before the touch is initiated [12]. The remainder of the cues were chosen based on a pilot survey, which asked 9 participants about 18 different cues that might be conveyed through touch. Those cues that people most wanted to use in touch communication (either as toucher or receiver) were included in the study.

The touchers were allowed to touch the receivers anywhere on the hand, arm or shoulder, and were told that they could perform any kind of touch or combination of touches that they felt was appropriate, and to take as long as they wanted. After the toucher signaled through keyboard press that they had finished, the receiver was presented with buttons on the screen labelled with the receiver choices (Table 1), and had to select one to indicate what they thought the toucher was communicating. If the receiver selected 'other', they were prompted to type in a response.

Additionally, on the very first presentation of each cue, the receiver was prompted to type in an open response with no forced choice cues, in order to get an indication of touch communication accuracy when totally naïve.

TABLE I. CUES FOR THE TOUCH COMMUNICATION TASK

| Cue label | toucher cue | receiver choices |
|---|---|---|
| attention | You just heard about something that your partner might find interesting. Try to get their ATTENTION through touch. | Your partner is trying to get your ATTENTION. |
| love | Think of all the wonderful qualities that your partner has, and how they enrich your life. Try to express LOVE through touch. | Your partner is trying to express LOVE. |
| happiness | You have just received good news. You are feeling very happy and you want to let your partner know. Try to express HAPPINESS through touch. | Your partner is trying to express HAPPINESS. |
| calming | Your partner is feeling upset thinking about a situation that cannot be changed. Try to be CALMING through touch. | Your partner is trying to be CALMING. |
| sadness | You have just received bad news. You are feeling very sad and you want to let your partner know. Try to express SADNESS through touch. | Your partner is trying to express SADNESS. |
| gratitude | Your partner has just helped you solve a problem. Try to communicate GRATITUDE through touch. | Your partner is trying to communicate GRATITUDE. |
| sympathy | Your partner is feeling hurt or upset and would appreciate emotional support. Try to communicate SYMPATHY through touch. | Your partner is trying to communicate SYMPATHY. |
| amusement | Your partner has just told a funny joke. Try to communicate AMUSEMENT through touch. | Your partner is trying to communicate AMUSEMENT. |
| other |  | Your partner is trying to communicate something else. |

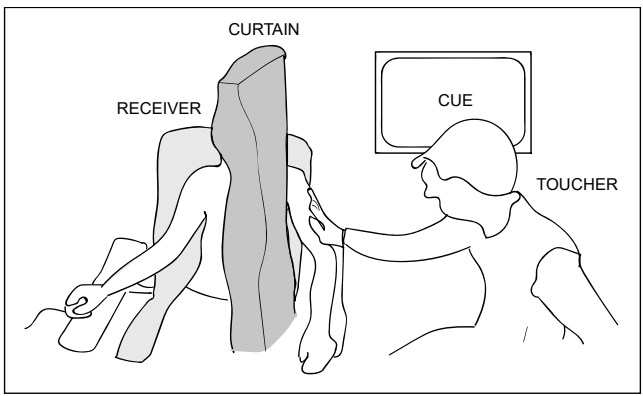

Figure 1. The experimental setup. The toucher was presented with a cue on a computer monitor, and then pressed a button on a keyboard to indicate the start and the end of each touch, controlling the pacing of the experiment. The receiver could then select their response using a mouse, and enter open-ended responses with a keyboard. The participants could not see each others' monitors or faces.

Both participants were instructed to remain quiet throughout the touch communication task, not talking or making any other noise such as laughing or sighing. The participants were not provided with any feedback.

The toucher cues were presented in a pseudo-random order, and the button labels that the receiver saw had their position shuffled on every trial. The cues were delivered and the responses were recorded using a custom python script using PsychoPy [13].

In the first half of the session with a pair, the receiver was totally naïve, and had no expectations. In the second half, after they swapped roles, the receiver had previously been the toucher, and was familiar with the contextual cues (Table 1; although we did not tell them they saw the same cues, and participants were not allowed to discuss the task, it was reasonable for them to assume so).

The prompt for an open-ended response on the very first presentation was presented only once per pair, in their initial roles only. Each cue was always presented four times: in their initial roles it was one open-ended response and 3 forced-choice responses; after switching roles it was four forced-choice responses.

*D. Data analysis*

Data were analyzed using R [14], and figures for these data were created using the packages ggplot2 [15] and patchwork [16]. The communication data were analyzed with a 1-sample proportions test against the chance performance rate. This also provided 95% confidence intervals around the overall proportion, which are non-symmetrical because the data are bounded. The datasets, analysis scripts and experiment code for the current study are available on the Open Science Framework [17].

To analyze the open-ended responses that participants typed in at the beginning of the experiment, we had three raters independently assign the label that they thought best matched what the participants had typed in. The raters were given the following instructions: *"You will see some text that was entered by participants as part of a touch experiment. Their task was to guess what another person was trying to communicate to them via touch. Your task is to choose the label that best matches their description."* They were presented with each typed response, one at a time, in a random order, above buttons on the screen labelled with the receiver choices (Table 1, but with the words "your partner is" replaced with "they were"). The raters did not know what cue the toucher had been trying to communicate, and if the typed response did not match any of the cues, they labelled it as 'other'. Some participants had entered their responses in Swedish, in which case they were translated to English before the raters made their judgments.

For each of the six cue words, we collected a total of 19 open-ended responses and 57 labels applied by the three raters. All 57 labels were included in all figures and analyses, i.e. all three labels for each single typed response were used. No attempt was made to evaluate how well the raters labelled the typed responses. Instead, the raters' labels were evaluated according to whether they matched the toucher's cue. The data were entered into 1-sample proportions tests, but the *n* was weighted by 1/3, in order to avoid inflating statistical power.

## III. RESULTS

*A. Touch communication effectiveness*

When making forced-choice responses, all receivers (including in both initial and swapped roles) interpreted the touches with an overall performance of 50% correct (95% CI = 46%, 54%), which was significantly greater than the chance rate of 12.5% ($\chi^2$ = 1341, n = 1056, p = 1.68e-293). All individual cues were also interpreted significantly better than chance (Fig. 2). The confusion matrix (Fig. 3, top) shows that all cues were labelled correctly more than they were mistaken for any other single cue. However, there were some consistent mistakes. Notably, calming and sympathy were often mistaken for each other, which may be due to their conceptual similarity. Happiness and amusement are also conceptually similar, and often mistaken for each other, although both were also misinterpreted as attention.

We also examined the open-ended responses from the beginning of the experiment, in which the initial receivers had no experience of any of the touches. The confusion matrix (Fig. 3, bottom) shows the raters' labelling of the participant responses according to which cue was presented to the toucher. Despite the near complete lack of any context, the raters' labels matched the toucher's cue surprisingly well for attention (95% CI = 23%, 80%), love (19%, 75%) and calming (10%, 64%). In this case it is not clear what chance performance should be, since there are an infinite number of possibilities for the receiver (a fact which is reflected in the relatively large number of "other" labels applied to the responses). However, both attention and love were correctly identified at rates significantly better than the chance rate (12.5%) for the forced-choice task (attention: $\chi^2$ = 24.4, p = 7.71e-07; love: $\chi^2$ = 16.1, p = 5.88e-5; n weighted to equal 19).

Agreement between the raters in how they labelled the responses was high. We found that for 81% of responses, at least two of the three raters agreed on the label (between 13 and 18 of the 19 responses to each cue). Some examples of responses for which at least two raters' labels matched the toucher's cue include "playfulness, games" for amusement, "thanking or greeting" for gratitude, and "love" for love.

Some examples of responses for which none of the raters' labels matched the toucher's cue include "excitement" for attention (labelled happiness), "caring for someone, love" for calming (labelled love), and "safe" for sadness (labelled calming). Examples of responses that were labelled as "other" by at least two raters include "fear", "anger" and "irritation".

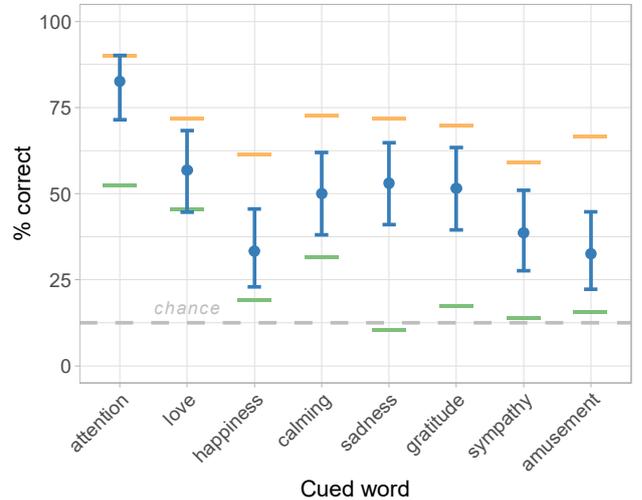

Figure 2. Touch communication effectiveness (blue), showing the overall percentage correct with 95% confidence intervals around the percentage (non-symmetrical since data are bounded). Data shown are from all pairs from both halves of the session, forced-choice responses only (132 presentations of each cue). All cues were correctly interpreted significantly better than chance (12.5%, dashed line) using one-sample proportions tests and Bonferroni correction for multiple comparisons (family-wise α = 0.05). Green markers indicate performance based on open-ended responses that were later labelled by naïve raters ('correct' here means that the label matched the toucher's cue). Gold markers indicate the discriminability of the touch cues, calculated from data in which each receivers' most common response to each cue was treated as correct.

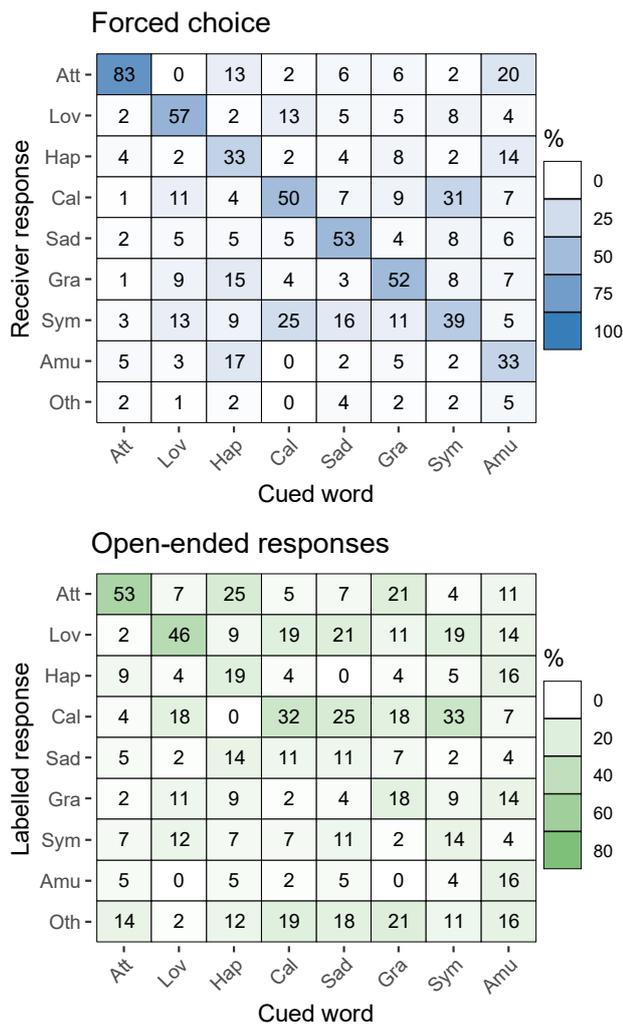

Figure 3. Confusion matrices for the touch communication task (columns may not sum to exactly 100 because individual cells have been rounded). **Top:** Data from forced-choice responses. The receiver chose from a list of options to indicate what they thought was communicated (132 responses for each cue). **Bottom:** Data from open-ended responses. In their initial roles, the first time the receiver had experienced any touches in the experiment, they had to type in what they thought was being communicated to them. Naïve raters then independently labelled the cue that best matched their responses (57 labels for each cue, applied to 19 responses).

Comparing performance on the forced-choice task when participants were in their initial roles vs when they swapped roles gives us some insight into how a small amount of additional context can help improve touch communication. Overall accuracy with initial roles was 41%, and with swapped roles was 56%. This was a significant improvement of 15% (95% CI = (9%, 21%), $\chi^2$ = 21.9, initial n = 456, swapped n = 600, p = 2.92e-06).

If the receivers (in either initial or swapped roles) had gotten feedback, they might have performed even better, with the limiting factor being how discriminable and consistent the touchers' chosen gestures were for each cue. We estimated the discriminability of the touch gestures by taking the response that each receiver chose most often for each cue and treating it as correct. These values are shown in Fig. 2 (gold markers).

### B. Relationship type

Participants were asked to define the nature of their relationship, which was categorized as romantic, friendship or family. We found no significant difference in overall performance according to relationship type ($F(2,16)$ = 0.8, $p$ = 0.453, *CI*: ±0.6). When analyzing the different cues separately, we found that gratitude was significantly better communicated within friendship pairs than within romantic pairs (difference = 31%, CI = (6%, 57%), $\chi^2$ = 10.3, romantic n = 62, friendship n = 56, p = 1.34e-03; family relationships were excluded due to low sample size).

### A. Relationship closeness

Participants rated the relationship with their partner for closeness, on a scale of 1 – 4. Because of the small range in scores, we treated closeness as a categorical variable. Overall touch communication performance was not significantly affected by the closeness score provided by the receiver (F(3,33) = 1.9, p = 0.157, CI: ±0.6), nor the toucher (F(3,33) = 2.9, p = 0.051, CI: ±0.6).

The differences in closeness scores within pairs ranged from 0 to 1, and provided an indication of how similarly the individuals within a pair rated the closeness of their relationship. Closeness difference scores did not significantly affect touch communication performance (F(2,15) = 1.0, p = 0.391, CI: ±0.6).

### B. Comfort with social touch

The participants answered the 'Social Touch Questionnaire' (STQ) which assesses the personal attitude towards social situations involving touch [18]. The STQ has a minimum score of 0, indicating relative comfort with touch, and a maximum of 80, indicating relative discomfort with touch. The overall mean score was 30.6 (SD = 11.4). STQ scores did not significantly differ between our female (M = 30.5, SD = 12.8) and male (M = 30.6, SD = 10.8) participants (difference CI = (-8.1, 7.8), t(36) = 0.04, p = 0.972).

Our observed social touch questionnaire scores were closer to the 'high social anxiety' group (M = 32.1) than the 'low social anxiety' group (M = 20.1) reported in [18]. In that study, the groups were chosen from a pool of American female undergraduates, selecting those with the top and bottom 25% of scores on the Social Phobia and Anxiety Inventory. The higher scores we observed may reflect a somewhat high level of social anxiety in our sample. It is possible that cultural differences may account for the relatively high scores, since we recruited from a Swedish university. Although the university is an international community, we did not collect any data on the cultural background of our participants.

We found no significant correlation between touch communication performance and STQ scores of the receiver (r = -0.15, CI = (-0.45, 0.18), p = 0.372), or of the toucher (r = -0.22, CI = (-0.50, 0.12), p = 0.185). The difference in STQ scores within pairs provides an indication of the similarity of the individuals within a pair in terms of their comfort with social touch, and ranged from 0 to 30 (median = 9). There was no significant correlation between STQ score difference and touch communication performance (r = 0.03, CI = (-0.43, 0.47); p = 0.907).

*C. Participant feedback*

Participants were asked to provide feedback on the touch communication task. For each cue, they were asked 'In general, how much do you want to communicate CUE to other people through touch?' and 'In general, how much do you want others to communicate CUE to you through touch?'. Answers were given on a visual analog scale with 'not at all' and 'very much' marked at the ends of the scale. Participants were also prompted to type in anything else they would like to communicate through touch.

Participants rated love and calming as the words they most want to communicate through touch; while attention and sadness were among those that participants least wanted to communicate. The cue had a significant main effect on VAS rating ($F(7,502) = 14.0$, $p = 2.00e-16$, CI: ±4.9). The role (whether they were asked about communicating or receiving the cue) had no significant effect on VAS rating ($F(1,592) = 0.09$, $p = 0.765$, CI: ±4.9, and there was no significant interaction ($F(7,592) = 0.39$, $p = 0.907$, CI: ±6.8).

IV. DISCUSSION

Our results show that adults in close personal relationships are capable of communicating through touch messages that vary in their emotional content. We found no big differences between the different types of relationships, individuals' comfort with touch, or other demographic and questionnaire data in our recruited pairs. It is possible that the difference in communication performance reported in [11] can be attributed to a difference in willingness to touch that has been observed between strangers and romantic couples [3], rather than that romantic couples develop an idiosyncratic touch language that is more effective.

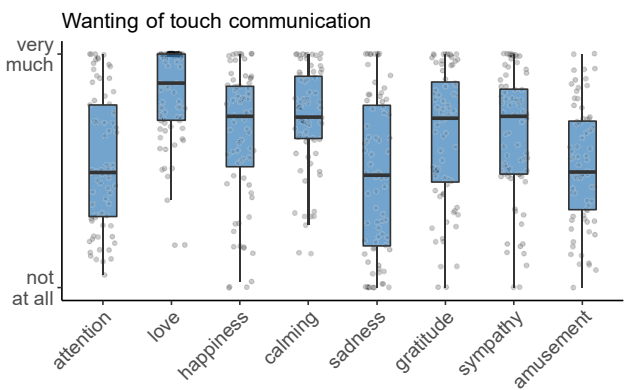

Figure 4. For each cue, participants were asked 'In general, how much do you want to communicate CUE to other people through touch?' and 'In general, how much do you want others to communicate CUE to you through touch?'. Answers were given by clicking on a location along a line – a visual analog scale – with 'not at all' and 'very much' marked at the ends of the line. Responses to the two questions are combined.

Consistent with this, we found that the receivers' responses to the same cues were not perfectly consistent, suggesting that the touch strategies used by the touchers were not perfectly distinguishable. It is possible that training people to use optimal touch strategies could provide more accurate touch communication.

We also found that the use of a forced-choice paradigm where participants choose from a list of possible options provides a big boost to performance. The existence of an 'other' option does not adequately account for this effect, and was rarely chosen by participants in the forced-choice task. We also found that when receivers had already experienced the toucher role and seen the more detailed contextual cues, communication performance was improved. The experience of being receivers first could also have provided some insights that enhanced their ability to perform touches that were easier to interpret. These results suggest that touch communication functions within a broader social context, and additional information is used to disambiguate touch messages when available.

The responses to the open-ended question reveal the expectations that people have about what sorts of things are likely to be communicated by touch. People were best at identifying attention, love and calming, and also often interpreted as calming those gestures that were used to communicate sadness, gratitude and sympathy. In the forced-choice task, the sympathy label was often applied to calming, sadness and gratitude gestures, which may be a similar misinterpretation given the conceptual similarity and mutual confusability of calming and sympathy. These apparent expectations about what touch is used to communicate matches quite well with the finding reported in [10] that people consider touch the preferred channel for communicating love and sympathy.

Interestingly, communication performance contrasted somewhat with what people told us they wanted to use touch communication for. While love and calming had high 'wanting' ratings, attention was relatively undesirable. It is possible that participants' wanting ratings were simply associated with the emotional valence of the cue word, rather than on how effective or useful touch is for conveying that word.

ACKNOWLEDGMENT

We thank Mattias Savallampi, Morgan Frost Karlsson and Christos Korres for rating the open-ended responses. We thank Ann-Charlotte Johansson for translating responses entered in Swedish.